\documentclass[twocolumn,prl,aps]{revtex4}

\usepackage{graphicx}
\usepackage{dcolumn}
\usepackage{bm}

\begin{document}

\preprint{APS/123-QED}

\title{Probing the superfluid velocity with a superconducting tip\,: \\ the Doppler shift effect}

\author{A. Kohen$^1$, Th. Proslier$^1$, T. Cren$^1$, Y. Noat$^1$, W. Sacks$^1$, H. Berger$^2$ and D. Roditchev$^1$}
\affiliation{$^1$Institut des Nanosciences de Paris, I.N.S.P., Universit\'es
Paris 6 et 7, C.N.R.S. (UMR 75\ 88),  75015 Paris, France}
\affiliation{$^2$Institute of Physics of Complex Matter, E.P.F.L., 1015 Lausanne, Switzerland}%

\date{\today}

\begin{abstract}
We address the question of probing the supercurrents in
superconducting (SC) samples on a local scale by performing
Scanning Tunneling Spectroscopy (STS) experiments with a SC tip.
In this configuration, we show that the tunneling conductance is
highly sensitive to the Doppler shift term in the SC quasiparticle
spectrum of the sample, thus allowing the local study of the
superfluid velocity. Intrinsic screening currents, such as those
surrounding the vortex cores in a type II SC in a magnetic field,
are directly probed. With Nb tips, the STS mapping of the
vortices, in single crystal 2$H$-NbSe$_2$, reveals {\it both} the
vortex cores, on the scale of the SC coherence length $\xi$, and
the supercurrents, on the scale of the London penetration length
$\lambda$. A subtle interplay between the SC pair potential and
the supercurrents at the vortex edge is observed. Our results open
interesting prospects for the study of screening currents in any
superconductor.
\end{abstract}

\maketitle

 A fundamental property of the superconducting (SC) state is
its response to an applied magnetic field. In particular, the
field penetrates type II superconductors in the form of quantized
flux, or vortices, each one carrying a flux quantum of
$\phi_0$=h/2e, which are arranged in a lattice \cite{Abrikosov}.
Each vortex is surrounded by screening currents which decay over
$\lambda$, the magnetic penetration length, and has a core
extending over the coherence length $\xi$. The SC pair potential,
$\Delta$({\bf r}), decays from its maximal value outside the core,
down to zero in its center. As shown by Bardeen, Cooper \&
Schrieffer (BCS), in the spatially homogenous case the SC state
has an excitation spectrum given by $E_{\bf k} = (\varepsilon_{\bf
k}^{2}$+$\Delta^2)^{1/2}$. This results in a unique quasiparticle
density of states (DOS) in which a gap of width $\Delta$, with a
peak at its edge, opens at the Fermi level. In the vortex state,
the DOS in the SC becomes spatially inhomogeneous due to both the
currents flowing around the vortices and the variations in
$\Delta$({\bf r}). The two different effects are expected to occur
on the length scales $\lambda$ and $\xi$ respectively. As can be
seen from the BCS spectrum, a consequence of changing
$\Delta$({\bf r}) is a modification of the DOS gap width. In addition,
as shown by Caroli et al.\cite{Caroli}, bound states are formed in
the vortex core since $\Delta$({\bf r}) acts as a potential well.
The bound states affect the low energy DOS and are significant
close to the vortex center.

The Scanning Tunneling Microscope (STM) is an instrument of choice
to map the DOS variations on a nanometer scale. The technique is
based on the tunneling current ($I$), flowing between a normal
metal tip and a sample, measured as a function of the tip position
and the bias voltage ($V$). Combined with spectro-scopy (STS), the
conductance $dI/dV({\bf r},V)$ reflects the sample local DOS,
which has been exploited to study the vortex lattice in several
materials
\cite{Hess,Hesslambda,Renner,DeWilde,Sakata,Davis,Eskildsen}. The
main focus of these experiments was a detailed study of the vortex
core bound states and/or the measurement of $\xi$, as inferred
from the spectra due to the spatial variation of $\Delta$({\bf
r}). Study of the screening currents, and thus measuring
$\lambda$, had proven to be more delicate.

In general when an uniform current flows in a SC sample, the
excitation spectrum can be rewritten as\,:
\begin{equation}
E_{\bf k}\, =\,(\varepsilon_{\bf
k}^2+\Delta^2)^{1/2}+m\mathbf{v_F} \cdot \mathbf{v_s} \ \ \ ,
\end{equation}
where $\mathbf{v_F}$ is the Fermi velocity and $\mathbf{v_s}$ is
the superfluid one \cite{Fulde}.
As long as the Doppler energy, $m\mathbf{v_F} \cdot
\mathbf{v_s}$, is small with respect to $\Delta$, the main effect
on the DOS is a reduction in the peak height. However this effect
is relatively small, only a few percent in magnitude, and is
therefore not effective, with STM, as a means to study the supercurrents.
While other methods are sensitive to magnetic field variations,
such as Hall probe or SQUID microscopies \cite{kirtley}, and are able to measure
$\lambda$, they lack the high spatial resolution of STM.
Moreover these
techniques are insensitive to $\Delta({\rm\bf r})$ and
thus cannot be used to determine $\xi$.

In pioneering STS experiments, Hess et al.\cite{Hesslambda} found that
even far from the core the spectra differed from the zero field
one, the main difference being a small in-gap shoulder. The
position of this shoulder is shifted to lower energies as one
approaches the core and finally merges with the peaks of the core
bound states. They interpreted this effect in terms of the Doppler
shift caused by the screening currents. However, as was
theoretically shown by \cite{Dahm}, this model is not
justified when the core states extend to large distances
and especially for low energies,
when they dominate the spectrum. Thus the the simple model based on the
Doppler shift, while giving the correct qualitative result, fails
to be quantative and should be replaced by a full solution of the
Eilenberger equations \cite{Dahm}.

Here we take a different approach and use a superconducting tip,
in STM/STS, to directly detect the pair supercurrents, due to
their Doppler shift effect on the quasiparticle spectrum. The use
of SC tips in low temperature STM was suggested \cite{Meservey}
and later realized by several groups
\cite{Pan,Suderow,Naaman,Giubileo,KohenC}. The potential
advantages of the SIS (SC-vacuum-SC) configuration are an enhanced
spectroscopic energy resolution and the possibility to
measure the Josephson pair tunneling \cite{Naaman}.
SC tips have so far been applied only at a single point;
no scanning spectroscopy has yet been reported. A priori, their
use to image the vortex lattice could result in complications.
First, the local magnetic field might affect the SC properties of
the tip and second, there is a force between the diamagnetic SC
tip and the vortices. A displacement of the vortices, leading to
a distortion of the STS images, is possible. In this letter we
report the successful use of superconducting Nb tips for STS
mapping of the vortex lattice in a NbSe$_2$ sample. Owing to the
enhanced spectroscopic resolution of the SC tip, we are able to
detect both the existence of the in-core bound states, and the
supercurrents flowing around the vortex. The latter is achieved by
the strong effect of the Doppler shift on the gap-edge peak
amplitude.

\begin{figure}[h]
\includegraphics[width=8.5cm]{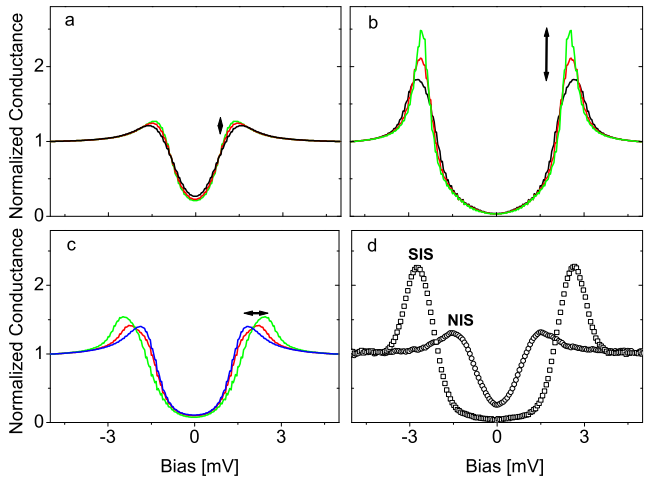}
\caption{\label{fig:epsart1} Calculations of\,: (a) normalized
tunneling conductance for a normal metal tip and a SC sample, with
$\Delta$ = 1 meV, T = 2.3 K, Doppler energy: 0, 0.3, 0.5 meV. (b)
conductance with a {\it superconducting} tip $\Delta_{tip}$ = 1.5
meV, and identical parameters as in (a). (c) SIS conductance with
large Doppler shift = 0.65 meV and varying surface gap
$\Delta_{sample}$ = 0, 0.3, 0.6 meV. (d) {Experimental} tunnel
conductance for a NbSe$_2$ sample with a SC Nb tip (SIS) and with
a normal Pt/Ir tip (NIS), T=2.3 K.}
\end{figure}

The effect of the supercurrents on a NIS spectrum, as calculated
using BCS, is shown in Fig.1a. One can see that if
m$\mathbf{v_F} \cdot \mathbf{v_s}<\Delta$, where Eq.(1) for the
Doppler shift holds, the corresponding change in the tunneling
conductance spectrum is very small and thus is difficult to observe
experimentally. However, in the SIS geometry (i.e. with a SC tip)
the same change in the sample DOS due to the Doppler shift leads
to a significant drop in the tunneling conductance peak amplitude,
as shown in Fig.1b. The enhancement of the effect is clearly due
to the overlap of the tip and sample DOS gap edges. Closer to the
vortex core, the principal change of the quasiparticle DOS should
be the decrease in the magnitude of $\Delta$, manifested by the
gradual shift of the tunneling conductance peaks towards lower
energies, see Fig.1c. Thus, in the SIS configuration, one would
expect two different behaviors in the SIS tunneling conductance
spectra while approaching the vortex core\,: the decrease in the
peak {\it amplitude} at large distances and the peak {\it shift}
to lower energies close to the vortex core. As we will
demonstrate, the corresponding length scales may be identified, in
a first approximation, as the penetration length $\lambda$ and the
coherence length $\xi$, respectively. The peak amplitude
variation, together with the Doppler shift energy, give the
profile of the supercurrent intensity as a function of distance
from the vortex center. By solving the London equation, a
quantitative fit yields the values of both $\xi$ and $\lambda$.

NbSe$_2$ crystals, grown using the standard Iodine Vapor Transport
technique \cite{Berger}, were studied using our home-made UHV STM
setup allowing a minimum temperature of 2 K and an applied
magnetic field up to 6\,T. Our low temperature STM unit,
described in \cite{Cren}, is digitally controlled and the
high-speed data acquisition allows a full spectroscopic mapping of
the sample. The preparation and characterization of our Nb tips is
reported in \cite{KohenC}. As a simple check, we first
measured the spectra of the NbSe$_2$ sample in zero field, and at
T = 2 K (see Fig.1d). These exhibit the typical features of a
SIS junction\,: sharp peaks appear at the voltages
$\pm$($\Delta_{tip}+\Delta_{sample})/e$. For comparison,
we show the spectrum using a normal Pt/Ir tip (NIS junction).
We thus obtain $\Delta_{Nb}$=1.5 meV and
$\Delta_{NbSe_2}$=1.0 meV, in agreement with their bulk values.

\begin{figure}[h]
\includegraphics[width=8 cm]{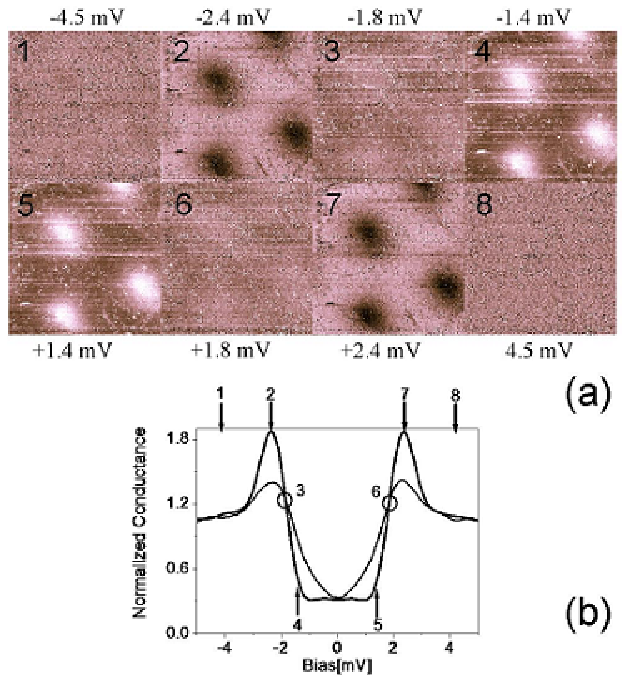}
\caption{\label{fig:epsart1} (a) Fixed scale conductance maps of
the vortex lattice,  applied field .06 T, 330 nm $\times$ 330 nm
area, T=4.5\,K. The maps (1-8) correspond to the bias voltages as
indicated in the figure. The strongest contrast is obtained at
V = $\pm(\Delta_{Nb}$+$\Delta_{NbSe_2})/e$,
maps 2 \& 7, and at V = $\pm\Delta_{Nb}/e$, maps 4 \& 5.
(b) Normalized conductance spectra measured in the vortex core
(NIS) and in between the vortices (SIS). Arrows mark the selected
voltages for the conductance maps in (a).}
\end{figure}

In Fig.2 we show a typical STS result of a 330$\times$330 nm$^2$
scan of the sample surface at T=4.5 K and in a magnetic field of
0.06 T. At each point, of a 256$\times$256 point topographic
image, a complete $I(V)$ tunneling spectrum was acquired in the
sample bias range from -10 mV to +10 mV. The conductance spectra,
$dI/dV(V)$, were directly derived from {\it raw} $I(V)$ data. In
Fig.2a we present 8 most significant conductance maps selected
from 256 measured. The maps are displayed in a {\it fixed} gray
scale without any additional contrast treatment. It is clear from
the maps n$^{os}$ 2, 4, 5 and 7 that the
familiar triangular vortex lattice is
successfully revealed by STS with a SC tip.
Here, the inter-vortex distance
of 190 nm matches the theoretical value $d = (2 \phi_0/ \sqrt{3}
B)^{1/2} \simeq$ 200 nm, for $B$ = 0.06 T. First, contrary to the
case of STS with a non-superconducting tip, the vortices do not
appear in the conductance maps near zero bias but rather at higher
bias values. The maximum contrast is achieved at
eV$\simeq$$\Delta_{tip}$ and
eV$\simeq$$\Delta_{tip}$+$\Delta_{sample}$. Second, one observes
an almost perfect symmetry of the contrast with respect to the
Fermi level. Indeed, the map n$^o$ 2 taken at -2.4 mV is almost
identical to n$^o$7 obtained at +2.4 mV. In these two maps the
vortices appear black due to the lower tunneling
conductance in the vortex cores. The maps n$^{os}$ 4 and 5, taken
at -1.4 mV and +1.4 mV respectively, are also quasi-identical, but
the vortices appear in white as the regions of higher
conductance.

The origin of the map contrast is better understood from the two
spectra plotted in Fig.2b. The first, obtained in the center of
the vortex, $\sigma_{NIS}(V)$, shows the characteristic NIS shape,
while the second, $\sigma_{SIS}(V)$, taken at a point in between
the vortices, shows the SIS one. At voltages above
($\Delta_{tip}$+$\Delta_{sample}$)/e, such as V$_8$, both spectra
have a common conductance value and no contrast is observed.
For lower $V$, a high peak (at V$_7$) develops in the SIS
spectrum at eV $\simeq \Delta_{tip}+\Delta_{sample}$ where it
exhibits a higher conductance than the NIS one. The two
spectra then cross at V$_6$. For V$<$V$_6$,
$\sigma_{SIS}$$<$$\sigma_{NIS}$ and a second high contrast can be
found for eV$\simeq$$\Delta_{tip}$, at V$_5$. Finally at V=0,
$\sigma_{SIS}$ $\simeq$ $\sigma_{NIS}$\,: the contrast is
negligible. We see why the contrast is inverted between maps 7 and
5 (respectively 2 and 4), and is small in map 6 (or 3).
Qualitatively, features in the sample DOS, besides being enhanced,
are shifted in energy by an amount $\Delta_{tip}$. By selecting
the sharpest positive and negative maps, we find the values $\pm
1.4$ mV and $\pm 2.4$ mV, respectively, and thus determine in a
different way\,: $\Delta_{Nb}$=1.4 meV
and $\Delta_{NbSe_2}$=1.0 meV.

To study the vortex bound states and the superfluid velocity
profile, we focused on a single vortex lattice unit cell and
reduced the temperature to T=2.3\,K, thus improving both spatial
and energy resolutions. The conductance maps at three selected
bias voltages (among 256) are shown in Fig.3. The apparent size
and shape of the vortices depend sensitively on the particular
bias\,: Map (a), obtained at the SIS spectral
peak, eV$\simeq$$\Delta_{tip}$+$\Delta_{sample}$, shows a much
larger diameter than map (c), obtained at
eV$\simeq$$\Delta_{tip}$, which reveals essentially the vortex
core. Such an enlargement is precisely due to the Doppler effect
of the screening currents. Map (b), obtained at the SIS/SIN
crossing-point voltage (V$_6$ in Fig. 2b) where little contrast
was expected, reveals a particular star shaped halo.

\begin{figure}[h]
\includegraphics[width=7.8cm]{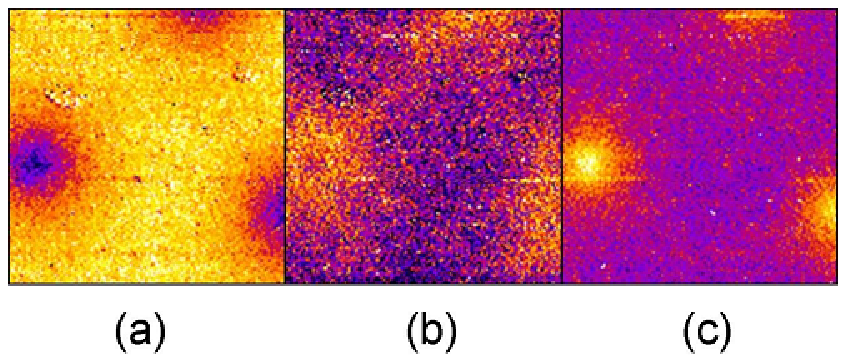}
\caption{\label{fig:epsart1} Conductance maps of a 210$\times$210
nm$^2$ area, T=2.3 K, applied field .05 T. Maps selected at
voltages: $V$=2.6 mV, 2.2 mV and 1.6 mV corresponding to (a) the
SIS peak, (b) the SIS/SIN intersection point and (c) the NIS peak.
These emphasize respectively\,: (a) the long range variations in
the SIS peak height, (b) the vortex star shaped halo and (c) the
vortex core. The scale of each map is readjusted to obtain the
best contrast, hence the apparent lower resolution in (b).}
\end{figure}

The dynamics of the spectral shape, for different distances from
the vortex center, is displayed in Fig.4. To reduce noise, each
spectrum is the average over a circle of radius $r$, concentric with
the vortex. One clearly sees the evolution from a SIS spectrum (A)
obtained far from the vortex, with a wide apparent gap and
pronounced peaks, to NIS spectra with low peaks and a narrower gap,
obtained in the vicinity of the core. At even smaller distances
($r \lesssim $10 nm) we observe a subtle dip - hump feature (D, H)
developing just above the gap edge, becoming more pronounced as we
approach the vortex center (B). There, a slight increase
of the amplitude of the peak is observed. The effect is
the signature of the vortex core states which exist near the
Fermi energy, but which are shifted to a voltage above
$\Delta_{tip}$/e in our case.

\begin{figure}[h]
\includegraphics[width=7.4cm]{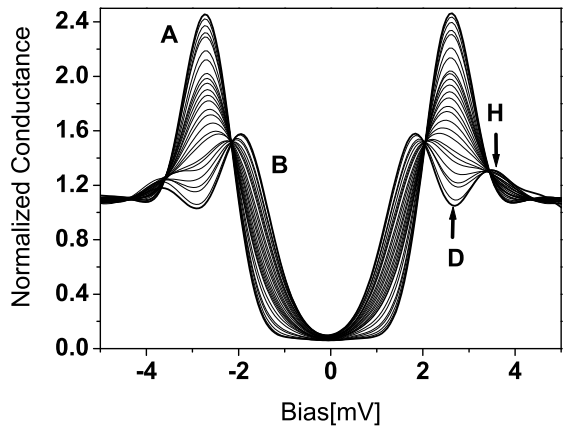}
\caption{\label{fig:epsart1} Evolution of the conductance spectra
as a function of distance from the vortex center. Far from the
vortex (A) the spectra show SIS features with high peaks at
$(\Delta_{tip}$+$\Delta_{sample})/e$. The peak height is first
lowered and then followed by a shift to a lower voltages, near
$\simeq \Delta_{tip}/e$. Close to the vortex center (B) a dip hump
feature appears (D, H), associated with a slight increase in the
peak.}
\end{figure}

\newpage
The spatial changes in the spectra, as a function of distance $r$
from the core, are summarized in Fig. 5 by the plots of the peak
position and amplitude.
Starting from a large distance, the peak height, initially at 2.4,
is continuously diminished, finally leveling off at a value $\sim$1.5
at $r = r_c \simeq 110$ \AA\, (dashed line).
This lowering of the peak amplitude (predicted in Fig.1b and
clearly visible in Fig. 4)
is due to the increase in the superfluid velocity as one approaches
the vortex core, which saturates at the critical velocity,
${\bf v} \simeq {\bf v}_c$ at $r \simeq r_c$.
On the contrary, the peak {\it energy}
is roughly constant (at $\sim$2.6 meV) for all $r \gtrsim 2 r_c$ but then
decreases rapidly ($r \sim r_c$) down
to its minimum value $\sim$1.8 meV at the vortex center. Thus, the peak
position as a function of $r$, shifted downwards by $\sim \Delta_{tip}$,
matches the pair potential profile.
As the mid-point of this profile is commonly used to estimate $\sqrt{2} \xi$,
we find directly $\xi \simeq 80$ \AA\, and from the peak amplitude profile,
$\lambda \simeq 750\,$\AA. In short, Fig.5 gives a picture of the pair potential
and supercurrent profiles of a vortex, of core radius $r_c$.

\begin{figure}[h]
\includegraphics[width=8.4cm]{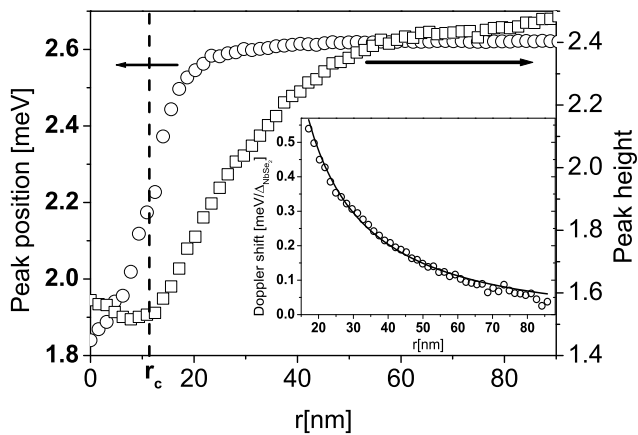}
\caption{\label{fig:epsart1} Gap-edge peak energy (circles) and
amplitude (squares) as a function of distance from the vortex
center.
Inset shows the Doppler shift energy,
in units of $\Delta_{NbSe_2}$, as a function of
distance from the vortex core (circles). The solid line is a
theoretical fit,
$\frac{\pi}{2}$\,$\frac{\xi}{\lambda}\,$K$_1$$(r/\lambda)$ with
the best fit obtained using $\lambda$=68 nm and $\xi$=6.6 nm.}
\end{figure}

The values of $\lambda$ and $\xi$ may be extracted in a different
way, through the Doppler shift energy (from Eq. 1) as a function
of $r$ (see inset of Fig.5).
We have fitted the data with the function\,:
$\frac{\pi}{2}$$\frac{\xi}{\lambda}$K$_1$$(r/\lambda)$, derived
from the London equation for an isolated vortex, where $K_1(x)$
is the modified Bessel function of order\,1. (This approximation
neglects Fermi surface and gap anisotropies.) The best fit is
found using $\xi$=66 \AA\, and $\lambda$=680 \AA\, and is shown as
the line in the inset. Finally, the Doppler shift at $r = \xi$,
together with the Fermi momentum $m v_F = \hbar k_F \simeq \hbar
\pi/2a$, where $a \simeq 3.5$ \AA\, is the lattice constant, leads
to $v_c \approx 180$ m s$^{-1}$ for the magnitude of the critical
velocity.

Our results, i.e. the vortex mapping using a SC tip, the detailed
evolution of the SIS to SIN spectra, and the Doppler shift energy
variation with distance from the core, all indicate that no
perturbation was observed,
arising from an interaction between the SC diamagnetic tip
and the vortices. Furthermore, the Nb tips remain
superconducting even under a field up to $\sim$ 0.3 T (the field
in the vortex core for an applied field of 0.05 T) a higher value
than the bulk critical field, 0.2 T. Our previous studies on the
effect of magnetic fields on our Nb tips, with a normal metal Au
sample, have shown that the tip critical field is enhanced up to 1
T (at T=4.2\,K). This result is due to the tip's apex size being
smaller than both $\xi$ and $\lambda$\, with the critical field
depending on the exact tip geometry \cite{KohenC}. In a small
number of experiments, currently under study,
we did observe a distorted vortex shape.

In conclusion, we have presented detailed STS mapping of the
vortex lattice using a superconducting tip, which opens new
possibilities for studying the velocity profile of the pair
currents, in any SC sample (type I included). We have demonstrated
that the features in the sample DOS are shifted in energy by an
amount $\Delta_{tip}$, as expected, but the principle result is
the direct effect of the Doppler shift on the SIS peak amplitude,
allowing a detailed mapping, on the nanometer scale, of the
currents in the superconductor. It could then be applied to cases
where there is an external current source, an oblique magnetic
field, or to confined superconductors, where giant vortices are
predicted. The presence of spontaneous local currents in high-Tc
superconductors, near the low-temperature pseudogap transition,
could be checked. Finally, we have demonstrated the stability of
our Nb tips for use in STS, paving the way for the future
measurement of the Josephson current.

{\small The authors thank F. Debontridder and F. Breton for
their technical assistance.
Sample preparation was supported by the NCCR
research pool MaNEP of the Swiss NSF.}


\end{document}